\documentclass[12pt]{article}
\usepackage{amsmath}
\usepackage{graphicx,psfrag,epsf}
\usepackage{enumerate}
\usepackage{natbib}
\bibpunct{[}{]}{,}{a}{}{;}
\usepackage{url} % not crucial - just used below for the URL 

\newcommand{\blind}{0}

\usepackage[colorlinks=true, citecolor=blue, linkcolor=blue]{hyperref} % comment out and use url package for submission

\usepackage{tikz}
\usetikzlibrary{shapes,backgrounds}

\begin{document}

\def\spacingset#1{\renewcommand{\baselinestretch}%
{#1}\small\normalsize} \spacingset{1}

%%%%%%%%%%%%%%%%%%%%%%%%%%%%%%%%%%%%%%%%%%%%%%%%%%%%%%%%%%%%%%%%%%%%%%%%%%%%%%

\if0\blind
{
  \title{\bf Embracing Data Science} %Understanding Data Science and Why We Should Teach It}
  \author{Adam Loy\\
    Department of Mathematics, Lawrence University}
    \date{August 26, 2015}
  \maketitle
} \fi

\if1\blind
{
  \bigskip
  \bigskip
  \bigskip
  \begin{center}
    {\LARGE\bf Embracing Data Science}
\end{center}
  \medskip
} \fi

\bigskip
\begin{abstract}
Statistics is running the risk of appearing irrelevant to today's undergraduate students.  Today's undergraduate students are familiar with data science projects and they judge statistics against what they have seen. Statistics, especially at the introductory level, should take inspiration from data science so that the discipline is not seen as somehow lesser than data science. This article provides a brief overview of data science, outlines ideas for how introductory courses could take inspiration from data science, and provides a reference to materials for developing stand alone data science courses.
\end{abstract}

\noindent%
%{\it Keywords:}  3 to 6 keywords, that do not appear in the title
%\vfill

 \spacingset{1.45} % DON'T change the spacing!

\section{Introduction}
\label{sec:intro}

Statistics is running the risk of appearing irrelevant to many of today's undergraduate students. At first, this may sound absurd since the number of statistics degrees awarded at the undergraduate level approximately tripled from 2003 to 2013 \citep{pierson2014}. I am not, however, referring to the popularity of the major, but rather the opportunities we as statistics educators have largely been missing. Rather than emphasizing that statistics is about ``thinking with and about data'' \citep[p. 3]{cobb2015}, we emphasize only part of the statistical thought process in our curricula. This issue is most pronounced at the introductory level. Consequently, we are failing to communicate what the field of statistics is about, making statistics seem irrelevant to many students, and perhaps even to many of our colleagues. In my experience many students who believe statistics is irrelevant are interested in thinking with and about data, so where do they turn?  Many are turning to data science.\footnote{I base this assessment on personal experience and observing the increasing popularity of data science training programs, such as the Coursera data science certificate run by Caffo, Leek, and Peng at Johns Hopkins and the Data Camp tutorials.}
%Statisticians emphasize that statistics is about ``thinking with and about data'' \citep[p. 3]{cobb2015}, but is this generally reflected in our undergraduate curricula, especially in our introductory courses where we attempt to provide a brief survey of the discipline? I  argue that the answer is no, though there are exceptions. Rather than emphasizing the entire process of thinking with and about data, our curricula often focus on individual pieces of this process or fail to communicate the overarching data analytic cycle (i.e., the entire thought process) to students. From my experience, it is this shortcoming that makes statistics seem irrelevant to many students, and perhaps even many of our colleagues. 
%Many students with this view, however, see value in data analysis and \emph{want} to learn from data, so where do they turn? It appears that many are turning to data science.\footnote{I base this assessment on personal experience and observing the increasing popularity of data science training programs, such as the Coursera data science certificate run by Caffo, Leek, and Peng at Johns Hopkins and the Data Camp tutorials.}

\section{What is Data Science?}

% So what is data science? While this question seems like it should have an easy answer, I have been at conferences where, even with only one discipline in the conversation, we have been unable to come to a consensus. This may in part be due to the popularity of the phrase in the media. At times it seems synonymous with applied statistics, while at other times it doesn't. 

In 2001, W. S. \citeauthor{cleveland2001} outlined a plan for the discipline of statistics. This plan encouraged statistics departments to focus on the practice of data analysis, resulting in an altered field called data science. If one takes Cleveland's view of data science, then it is a subset of statistics, essentially equivalent to applied statistics. Perhaps if every statistician, or at least a vast majority, had bought into Cleveland's plan for the discipline this would be the case, but not all statistics departments/curricula resemble Cleveland's ideal, so another definition has emerged.

\begin{figure}
\centering
\includegraphics{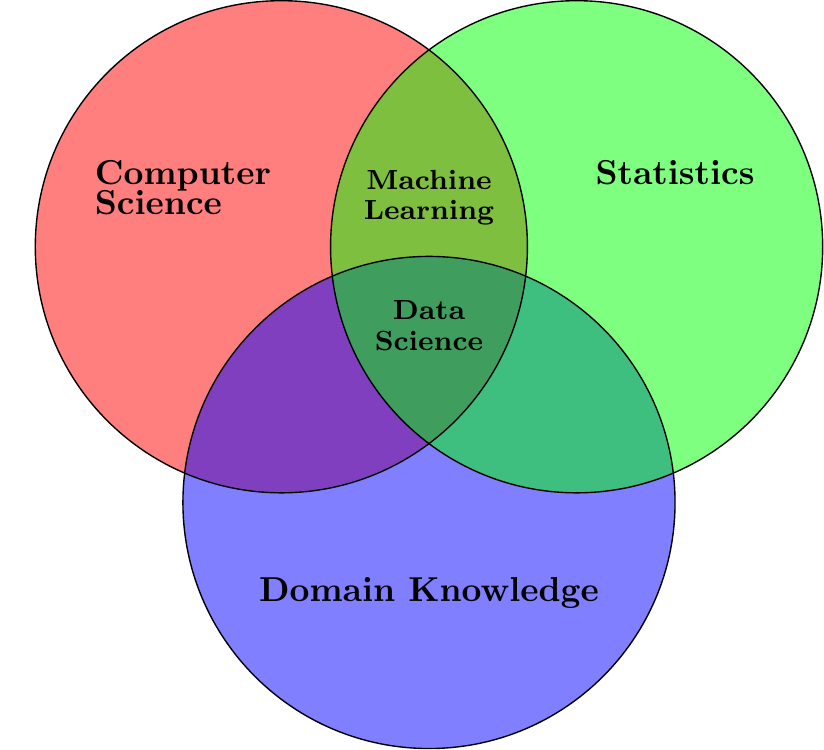}
\caption{\label{fig:venndiagram} An adaptation of Drew Conway's diagram explaining what knowledge and skills a data scientist needs. The original venn diagram was published on Drew Conway's blog in 2010.}
\end{figure}

Figure~\ref{fig:venndiagram} shows an adaptation of Drew Conway's venn diagram summarizing his view of data science \citep{conway}. According to Conway, data science is the intersection of statistics, computer science\footnote{Conway takes a narrower view, using the phrase ``hacking skills,'' but this view seems needlessly limiting.}, and domain knowledge. What is clear from figure~\ref{fig:venndiagram} is that all three areas are necessary to define data science---e.g., without domain knowledge, we would just be talking about machine learning. Consequently, we cannot view data science as simply a subset of statistics, but rather it \emph{utilizes a subset of statistics}. This intersection defining data science is what is often visible to undergraduate students in the media---for example, \emph{The Upshot} section of \emph{The New York Times} often presents data science products on a variety of topics. So it is important to understand data science in order to understand the background and expectations of our students.

Having defined what data science is, we must now turn to what a data scientist does. To better understand this, I have found the three steps of a data science project outlined by \cite{Wickham2014} to be a useful guide:
\begin{enumerate}
\item Collect data and questions.
\item Analyze the data, using both exploratory and confirmatory methods.
\item Communicate the results in a way that convinces someone to ``take action.''
\end{enumerate}
At first, the above steps may sound identical to the steps in a applied statistics project, but I find statisticians and data scientists apply different weights to the tasks outlined above. For example, many statisticians downplay the importance of graphics in formal analyses believing that they are not rigorous enough, while data scientists fully embrace the power of graphical exploration. Further, many statisticians focus on formally reporting the results of their analyses, often in ways that are not accessible to a general audience. While the best applied statisticians are able to communicate their findings to a general audience, this seems to be a core requirement of a data science. Data scientists take communication to another level through the development of web apps and other data products, with the goal of appealing to a general audience.

%Some statisticians will not perceive a difference in the above steps and what they perceive statistics to be; however, as a discipline statistics has not fully embraced all aspects of a data science project, especially in academia. Data scientists fully embrace the importance of graphics and data manipulation, but this is not the case in statistics. While many statisticians understand the importance of good graphics and tools for data manipulation, it is still common for this to be looked upon as ``lesser'' research. Until the field of statistics fully embraces all aspects of the data analytic cycle (CITATION??), there is a need to differentiate between statistics and data science.  
While I have focused on the differences between the statistics and data science, there are far more similarities than differences as both disciplines are deeply rooted in the data analytic cycle. So why not take inspiration from data science when we design our statistics courses?

%It is obvious that The above steps reveal that data science projects and statistical analyses are very similar. This similarity is not surprising since both statisticians and data scientists think with and about data. So why not take inspiration from data science when we design our statistics courses?

\section{How it can inform our teaching}

Being mindful that today's undergraduate students are familiar with data science projects and that they judge statistics against what they have seen will help us better teach statistics. Many students believe that statistics is irrelevant because we are not ``rising to the level'' of the data science projects they have seen. This is especially pronounced at the introductory level\footnote{I consider the ``traditional'' introductory statistics course to be one that closely follows the AP curriculum \cite{AP2010}. This includes the new randomization-based approaches to the course, e.g. \cite{lock2013} and \cite{tintle2015}.} because we still mainly focus on the traditional pillars of statistics---data collection via surveys and experiments, and inference---rather than focusing on the entire data analytic cycle. Since both statisticians and data scientists think with and about data, we should take advantage of our students' prior exposure to data science in our statistics curricula.
In this section I outline a way to incorporate the elements of a data science project into an introductory statistics course. Such an introductory course should provide a trickle down benefit to the upper-level courses, assuming that students take an introductory course prior. I also provide a reference concerning the development of a stand alone data science course.

\subsection{Introductory courses} 

The goals of an introductory course in any discipline are to (1) provide an overview of that discipline and (2) begin to develop a core knowledge base and skill set that is necessary for more advanced topics. I do not believe that many of the ``traditional'' introductory statistics courses achieve both goals, including those that I have taught. 
In fact, I think we often fail to achieve either goal because we try to cover too many modes of inference and some of us still emphasize by-hand calculations. 

Rather than trying to cover so many inferential procedures we should take inspiration from data science and strive to provide an overview of the entire data analytic cycle. Focusing on the entire cycle will reveal the core knowledge base and skill set that should be emphasized. More specifically, by breaking down the steps of a data science project the necessary topics that are largely missing from the ``traditional'' introductory course emerge. These topics are core aspects of thinking with and about data, so we must incorporate them into our courses to stay relevant.

\begin{enumerate}
	\item Collection
		\begin{enumerate}
			\item Collect and refine questions\\
			Too little time is spent questioning the data sets we use in class. While time constraints may be cited as a reason to discuss only ``focused'' analyses, a guided discussion of what questions could be answered using a specific data set will hone our students' thought processes. This will allow them to question data sets outside of our classes, rather than only being able to rely on a data analytic ``script.'' Based on the recommendations from the American Statistical Association \citep{ASA2005, ASA2014}, this change is already taking place, but we need to continue in this direction. One way to facilitate such discussions is to focus on case studies rather than textbook examples. %Such examples provide a natural space in class for a discussion where questions about the data set can be honed. 

			\item Collect data relevant to the questions\\
			Too often we provide students with the data directly, rather than having them collect or access data. In some courses, final projects provide students with their first opportunity to collect data. Rather than waiting until the end of the course, integrating data collection into the cycle will provide students with important tools for future analyses. In addition to traditional modes of data collection, it is essential to include some introduction to accessing web-based data.
		\end{enumerate}
	\item Analysis
		\begin{enumerate}
			\item Data manipulation\\
			We must stop ignoring data manipulation. Data manipulation is often the most time consuming task in the data analysis cycle, so how can we justify it receiving little or no coverage in our introductory course? Discussing tidy data \citep{wickham2014tidy} and the tools available to reshape data into the forms necessary for analyses are necessary additions %to build these fundamental skills.
to introductory courses. 

			\item Visualization\\
			Rather than going through a laundry list of exploratory graphics without motivation, we should incorporate graphics throughout the entire course. %Integrating the discussion throughout the course 
This will allow new graphics to be introduced on a just-in-time basis, and will avoid students thinking that graphics are unimportant or overly simplistic, as they seem to when we force the entire section into the first third of the traditional course. Further, we should refine the vocabulary we use in class by appealing to the grammar of graphics \citep{wilkinson2006grammar} and incorporate multivariate topics through the use of aesthetics and facetting. \cite{kaplan2015} provides an example of how this might be done at an appropriate level.

			\item Modeling\\
			We already discuss statistical models in introductory courses, but why must we treat $t$-tests and simple linear regression models separately? A unified discussion of modeling from a linear model perspective, such as the one taken by \cite{kaplan2012}, would help students focus on the thought process rather than the differences between formulas. Further, we need to discuss the inherent link between visualization and modeling in the data analytic cycle. This can easily be done if we model the thought process for our students during class and through assignments.

		\end{enumerate}
	\item Communication\\
	Many introductory statistics courses address communication in some way. Courses that require final projects seem to best develop a student's ability to communicate statistical results. However, we need to do more. Students are used to absorbing material online, so enabling them to communicate their findings online will have immediate impact. This can be done, for example, by teaching students how to build webpages using rmarkdown \citep{rmarkdown} and knitr \citep{Xie2013}, or basic web apps using Shiny \citep{shiny}. These skills can be built incrementally at the end of each case study, and will provide very marketable skills, as well as the basic skills to perform reproducible research.

%	Incorporating web-based presentation (in addition to traditional reports and presentations) into an introductory course will add value for our students. They are used to absorbing material online, so enabling them to communicate their findings online
	
%	 the students would immediately see the value we have added. For example, we could teach students how to build webpages using Rmarkdown \citep{rmarkdown} and knitr \citep{Xie2013}, or a basic web app using Shiny \citep{shiny}. These skills could be built incrementally at the end of each case study, and would provide very marketable skills, as well as the basic skills to perform reproducible research.

\end{enumerate}

While adapting an introductory statistics course to include these topics requires a lot of work, and numerous iterations to hone, the effort will be rewarded. After leaving a course where the entire data analytic cycle is examined I believe that students will be able to access new data sets and perform their own analyses in new situations. While less inferential material may be covered, students will understand how to manipulate and visualize data far better. (Many of the questions I field from former students relate to these topics, so they are certainly topics students encounter outside of our classes.) Additionally, less time will need to be devoted to developing basic data skills in later courses, adding room in the curriculum for more statistical topics. Finally, this approach provides a realistic overview of the field of applied statistics, and will help broadcast what statistics is about, even to those students we only see in an introductory course.

\subsection{Teaching data science}

In addition to changes in the introductory curriculum, the increasing visibility of data science opens the door for new courses. One possibility is to create a stand alone data science course. If there is not room in the curriculum for such a course, a hybrid course discussing more advanced statistical models and modern methods in a data science framework may be more feasible. I refer the interested reader to \cite{hardin2014} for details from seven varieties of data science courses.

% \section{How might we accomplish this?}

\subsection{Conclusion}

While statistics still seems to be increasing in popularity based on the number of degress granted, it is running the risk of appearing irrelevant to a large population of students. If we take inspiration from data science and focus attention on the aspects of a data science project in our applied statistics courses, then we will show students that statistics is relevant while providing them with the tools to think with and about data, starting in the introductory course. This can only strengthen our applied statistics curricula and draw students into the discipline.

\bibliographystyle{chicago}
\bibliography{datascibib}

%\section{Appendix}
%
%By a ``traditional'' introductory statistics course, I am describing a course that covers much of the following: 
%% 
%\begin{itemize}
%	\item Vocabulary for describing data
%	\item Displaying/describing univariate and bivariate distributions for quantitative and categorical variables.
%	\item Measures of center and spread for a quantitative variable
%	\item Simple linear regression and correlation
%	\item Data acquisition: an overview of sample surveys and experimental design
%	\item Basic probability rules
%	\item Understanding sampling distributions
%	\item Hypothesis testing and confidence intervals for: one proportion, a difference in proportions, one mean, a difference in means, a paired difference in means, and one-way ANOVA
%	\item Chi-square tests
%\end{itemize}
%% 

\if0\blind
{
\section{Acknowledgements}
I would like to thank the participants of the Harnessing Big Data Workshop sponsored by the Associated Colleges of the Midwest. The conversations we had there made me think more specifically about my views on data science and its role in the undergraduate curriculum.

%\section{About the Author}
%\begin{minipage}{0.3\textwidth}
%\includegraphics{AdamLoy-7-Edit.jpg}
%\end{minipage}
%\begin{minipage}{0.7\textwidth}
%Adam Loy is an Assistant Professor of Statistics at Lawrence University, where he has taught the entire statistics curriculum since 2013. Adam received his B.A. in Mathematics/Statistics from Luther College and his M.S. and Ph.D. in Statistics from Iowa State University. His research interests include statistical graphics, R development, linear mixed-effects models, and statistics education.
%\end{minipage}

} \fi
\end{document}